\def\nk{n_{\rm b}}

\def\rfr#1{Eq.~(\ref{#1})}
\def\rfrs#1#2{Eqs.~(\ref{#1})-(\ref{#2})}
\def\Rfr#1{Eq.~(\ref{#1})}

\def\derp#1#2{\rp{\partial{#1}}{\partial{#2}}}
\def\dert#1#2{\frac{{{d}}{#1}}{{{d}}{#2}}}

\def\virg#1{``#1"}

\def\eqi{\begin{equation}}
\def\eqf{\end{equation}}
\def\eqia{\begin{eqnarray}}
\def\eqfa{\end{eqnarray}}
\def\Om{\mathit{\Omega}}
\def\rp#1#2{{#1\over#2}}
\def\lb#1{\label{#1}}

\def\ton#1{\left(#1\right)}
\def\qua#1{\left[#1\right]}
\def\grf#1{\left\{#1\right\}}
\def\ang#1{\left\langle #1\right\rangle}

\RequirePackage{fix-cm}
\documentclass[smallextended,useAMS]{svjour3}       
\smartqed  
\usepackage{graphicx,amsmath,amscd,amssymb,latexsym}

\RequirePackage{color}
\journalname{Eur. Phys. J. C}
\begin{document}

\title{A comment on \virg{A test of general relativity using the LARES and LAGEOS satellites and a GRACE Earth gravity model}, by I. Ciufolini et al.
}


\titlerunning{A comment on \virg{A test of general relativity [\ldots]}}        

\author{Lorenzo Iorio
}

\authorrunning{L. Iorio} 

\institute{L. Iorio \at
              Ministero dell'Istruzione, dell'Universit\`{a} e della
Ricerca (M.I.U.R.), Viale Unit\`{a} di Italia 68
Bari, (BA) 70125,
Italy \\
              Tel.: +39-329-2399167\\
              \email{lorenzo.iorio@libero.it}           
}

\date{Received: 14 April 2016 / Accepted: 8 January 2017}

\maketitle

\begin{abstract}
Recently, Ciufolini et al. reported on a test of the general relativistic gravitomagnetic Lense-Thirring effect by analyzing about 3.5 years of laser ranging data to the LAGEOS, LAGEOS II, LARES geodetic satellites orbiting the Earth. By using the GRACE-based GGM05S Earth's global gravity model and a linear combination of the nodes $\Om$ of the three satellites designed to remove the impact of \textcolor{black}{errors in} the first two even zonal harmonic coefficients $J_2,~J_4$ of the multipolar expansion of the Newtonian part of the Earth's gravitational potential, they claimed an overall accuracy of $\textcolor{black}{5}\%$ \textcolor{black}{for the Lense-Thirring caused node motion}. We show that  the scatter in the nominal values of the uncancelled even zonals of degree $\ell = 6,~8,~\textcolor{black}{10}$ from some of the most recent global gravity models does not yet allow to reach unambiguously and univocally the expected $\approx 1\%$ level, being large up to \textcolor{black}{$\lesssim 15\%~(\ell=6),~6\%~(\ell=8),~36\%~(\ell=10)$} for some pairs of models.

\keywords{Experimental studies of gravity \and Experimental tests of gravitational theories \and Satellite orbits \and Harmonics of the gravity potential field }
\PACS{04.80.–y \and 04.80.Cc \and 91.10.Sp \and 91.10.Qm}
\end{abstract}

\section{Introduction}
In Ref.~\cite{2016EPJC...76..120C}, Ciufolini et al. reported a test of the general relativistic gravitomagnetic Lense-Thirring effect in the gravitational field of the Earth  using a linear combination of the right ascensions of the ascending nodes $\Om$ of the LARES, LAGEOS and LAGEOS II laser-ranged geodetic satellites and a GRACE-based Earth gravity model claiming an alleged overall systematic uncertainty of $5\%$.
For an overview of past attempts, see Refs.~\cite{2011Ap&SS.331..351I,2013OPhy...11..531R,2013NuPhS.243..180C} and references therein.

In this paper, we will quantitatively show that the claims in Ref.~\cite{2016EPJC...76..120C} are too optimistic despite  the recent advances in modeling the Earth's gravity field.

Let us recall that the gravitomagnetic node precession of a test particle in geodesic motion around a central rotating primary is \cite{LT18}
\eqi
\dot\Omega_\textrm{LT} \lb{LTtheo}  =\rp{2 GS}{c^2 a^3\ton{1-e^2}^{3/2}},
\eqf
where $G$ and $c$ are the Newtonian constant of gravitation and the speed of light in vacuum, respectively, $S$ is the body's angular momentum, while $a$ and $e$ are the orbit's semimajor axis and  eccentricity, respectively. \textcolor{black}{The magnitudes of the Lense-Thirring precessions for the satellites of the LAGEOS family, along with their relevant orbital parameters, are listed in Table \ref{tavola1}}.
\begin{table}
\caption{Relevant orbital parameters $a~(\textrm{km}),~e,~I~(\textrm{deg})$ of LAGEOS, LAGEOS II, LARES,  their classical node precession coefficients $\dot\Omega_{.\ell}$ for the first five even zonal harmonics ($\ell=2-10$), their Lense-Thirring node precessions $\dot\Omega_\textrm{LT}$ for $S = 5.86\times 10^{33}$ kg m$^2$ s$^{-1}$, and the combination coefficients $c_1,~c_2$. All the rates, in \textcolor{black}{milliarcseconds per year (mas yr$^{-1}$)}, are expressed to an accuracy of $0.1$ mas yr$^{-1}$ which corresponds to about a percent fraction of the predicted Lense-Thirring combined signature $\mathcal{C}_\textrm{LT}=50.2~\textrm{mas~yr}^{-1}$. The numerical values of the coefficients $c_1,~c_2$ are quoted to an accuracy adequate for a cancelation of $J_2$ to a sub-percent level.}
\label{tavola1}       
\begin{tabular}{llll}
\hline\noalign{\smallskip}
 & LAGEOS & LAGEOS II & LARES   \\
\noalign{\smallskip}\hline\noalign{\smallskip}
$a$  & $12270$ & $12163$ & $7828$ \\
$e$ & $0.0045$ & $0.0135$ & $0.0008$ \\
$I$  & $109.84$ & $52.64$ & $69.5$ \\
\noalign{\smallskip}\hline\noalign{\smallskip}
$\dot\Omega_{.2}$  & $4.159523197035\times 10^{11}$ & $-7.671024751108\times 10^{11}$ & $-2.0691803570443\times 10^{12}$ \\
$\dot\Omega_{.4}$  & $1.541082434098\times 10^{11}$ & $-5.57207688363\times 10^{10}$ & $-1.8385054326934\times 10^{12}$ \\
$\dot\Omega_{.6}$  & $3.29198354689\times 10^{10}$ & $4.98585219772\times 10^{10}$ & $-9.061255341802\times 10^{11}$\\
$\dot\Omega_{.8}$  & $2.3906795991\times 10^{9}$ & $1.10181009277\times 10^{10}$ & $-9.43157797573\times 10^{10}$\\
$\dot\Omega_{.10}$  & $-1.407631461\times 10^{9}$ & $-2.213156639\times 10^{9}$ & $3.04267201897\times 10^{11}$\\
\noalign{\smallskip}\hline\noalign{\smallskip}
$\dot\Omega_\textrm{LT}$  & $30.7$ & $31.5$ & $118.1$ \\
\noalign{\smallskip}\hline\noalign{\smallskip}
$c_j,~j=1,2$ & - & $0.344281069$ & $0.073388218$ \\
\noalign{\smallskip}\hline
\end{tabular}
\end{table}
\section{The systematic uncertainty due to the uncancelled even zonal harmonics}\lb{geopot}
One of the most critical issues of all the attempts planned or performed so far to detect the Lense-Thirring effect with Earth's artificial satellites is represented by the correct handling of the competing secular node precessions $\dot\Om_{J_\ell}$ induced by the even zonal harmonic coefficients\footnote{The parameters ${\overline{C}}_{\ell,0}$ \textcolor{black}{determining $J_\ell$} are the fully normalized Stokes coefficients of degree $\ell=2j,~j=1,2,3,\ldots$ and order $m=0$ of the geopotential's expansion. \textcolor{black}{For a given value of $\ell$, which can assume odd values as well, $m$ generally runs from 0 to $\ell$. The multipolar expansion of the gravitational potential includes also the Stokes coefficients ${\overline{S}}_{\ell,m}$; it is always ${\overline{S}}_{\ell,0}=0$, and the remaining ones do not induce secular orbital precessions \cite{Kaula00}. Thus, for a given degree $\ell$, there are $2\ell+1$ generally nonvanishing normalized Stokes coefficients.} } $J_\ell = -\sqrt{2\ell+1}~{\overline{C}}_{\ell,0},~\ell=2,4,6,\ldots$ of the multipolar expansion of the Newtonian terrestrial gravitational potential. Indeed, their nominal sizes are several orders of magnitude larger than the Lense-Thirring node precessions; moreover, their current level of modeling is not yet accurate enough to allow to use a single satellite's node to extract the relativistic signature. Thus, in the past years, some strategies to circumvent such an issue were devised. Contrary to what \textcolor{black}{is} claimed in Ref.~\cite{2016EPJC...76..120C}, the combination presumably
used in Ref.~\cite{2016EPJC...76..120C}, not displayed there, was explicitly proposed by the present author\footnote{See \cite{2011Ap&SS.331..351I,2013OPhy...11..531R} and references therein; in particular, \cite{2005NewA...10..616I}.} on the basis of a strategy proposed for the first time in Ref.~\cite{1996NCimA.109.1709C}. It was suitably designed to cancel out, at least in principle, the node precessions due to the first two even zonals $J_2,~J_4$, being \textcolor{black}{fully} impacted by all the other ones of higher degree $\ell\geq 6$. The hope is that, since the geopotential is modeled in the data reduction softwares used to process the satellites' data, the level of mismodeling in the uncancelled even zonals  is accurate enough to allow for a reliable determination of the Lense-Thirring effect with a percent accuracy.

Actually, as summarized in Refs.~\cite{2011Ap&SS.331..351I,2013OPhy...11..531R} and references therein, there are sound doubts that such a strategy could allow to reach a $\approx 1\%$ overall accuracy, as steadily claimed by Ciufolini et al. over the years. In this paper, we will enforce such concerns by showing that severe issues lurk even in the low-degree part of the spectrum of the geopotential.
\subsection{Review of the method used so far to extract the Lense-Thirring signature}\lb{methodo}
For the convenience of the reader, we will now review the aforementioned approach, which is a generalization of the one proposed for the first time in Ref.~\cite{1996NCimA.109.1709C}.

\textcolor{black}{By assuming to use the nodes of, say, $N$ different satellites, the following $N$ linear combinations can be written down
\eqi \mu_{\rm LT}\dot\Om_{\rm LT}^{(i)}+ \sum_{k=1}^{N-1}\ton{\derp{\dot\Om^{(i)}_{J_{2k}}}{J_{2k}}}\delta J_{2k},~i=1,2,\ldots N. \lb{lin} \eqf
They involve the Lense-Thirring node precessions as predicted by General Relativity (see \rfr{LTtheo}), scaled by a multiplicative parameter $\mu_\textrm{LT}$, and \textcolor{black}{errors in the computed}  secular node precessions due to  \textcolor{black}{errors in} the first $N-1$ even zonals $J_{2k},~k=1,2,\ldots N-1$, assumed as mismodeled through $\delta J_{2k},~k=1,2,\ldots N-1$.
In the following, we will use the shorthand \eqi\dot\Om_{.\ell}\doteq\derp{\dot\Om_{J_\ell}}{J_\ell}\eqf
for the partial derivative of the classical node precession with respect to the generic even zonal $J_\ell$ of degree $\ell$. Then, the $N$ combinations of \rfr{lin} are equated to the experimental residuals $\Delta\dot\Om^{(i)},~i=1,2,\ldots N$ of each node of the $N$ satellites considered. In principle, such residuals account for the purposely unmodelled Lense-Thirring effect, the mismodelling of the static and time-varying parts of the geopotential, and the non-gravitational forces. Thus, one gets
\eqi \Delta\dot\Om^{(i)} = \mu_{\rm LT}\dot\Om_{\rm LT}^{(i)}+ \sum_{k=1}^{N-1}\dot\Om^{(i)}_{.2k}\delta J_{2k},~i=1,2,\ldots N. \lb{lineq} \eqf
If we look at the Lense-Thirring scaling parameter $\mu_\textrm{LT}$ and the mismodeling in the even zonals $\delta J_{2k},~k=1,2,\ldots N-1$ as unknowns, we can interpret \rfr{lineq} as an inhomogenous linear system of $N$ algebraic equations in the $N$ unknowns
\eqi
\underbrace{\mu_{\rm LT},~\delta J_2,~\delta J_4 \ldots \delta J_{2(N-1)}}_{N},
\eqf
whose coefficients are \eqi\dot\Om^{(i)}_\textrm{LT},~\dot\Om^{(i)}_{.2k},~i=1,2,\ldots N,~k=1,2,\ldots N-1,\eqf while the constant terms are the $N$ node residuals
\eqi\Delta\dot\Om^{(i)},~i=1,2,\ldots N.\eqf
It turns out that, after some algebraic manipulations,  the dimensionless  Lense-Thirring scaling parameter, which is 1 in General Relativity, can be expressed as
\eqi\mu_\textrm{LT}=\rp{\mathcal{C}_\Delta}{\mathcal{C}_\textrm{LT}}.\lb{rs}\eqf
In \rfr{rs}, the combination of the $N$ node residuals
\eqi{\mathcal{C}}_\Delta \doteq \Delta\dot\Om^{(1)} + \sum_{j=1}^{N-1}c_j\Delta\dot\Om^{(j+1)} \lb{combo}\eqf
is, by construction, independent of the first $N-1$ even zonals, being impacted by the other ones of degree $\ell > 2(N-1)$ along with the non-gravitational perturbations and other possible orbital perturbations which cannot be reduced to the same formal expressions of the first $N-1$ even zonal rates. Instead,
\eqi\mathcal{C}_\textrm{LT} \doteq \dot\Om_{\rm LT}^{(1)} + \sum_{j=1}^{N-1}c_j\dot\Om_{\rm LT}^{(j+1)} \lb{comboLT}\eqf
combines the $N$ gravitomagnetic node precessions as predicted by General Relativity.
The dimensionless coefficients $c_j,\ j=1,2,\ldots N-1$ in \rfr{combo}-\rfr{comboLT} depend only on some of the orbital parameters of the $N$ satellites involved in such a way that, by construction, ${\mathcal{C}}_\Delta=0$ if \rfr{combo} is calculated by posing \eqi\Delta\dot\Om^{(i)}=\dot\Om^{(i)}_{.\ell}\delta J_{\ell},\ i=1,2,\ldots N\eqf for any of the first $N-1$ even zonals, independently of the value assumed for its uncertainty $\delta J_{\ell}$.}

In our case \cite{2005NewA...10..616I}, it is $N=3$; the satellites employed are LAGEOS,~LAGEOS~II,~LARES.
The resulting \textcolor{black}{combination} coefficients $c_1,~c_2$ yielding $\mu_\textrm{LT}$, worked out from \cite{Capde05}
\begin{align}
\dot\Omega_{.2} \lb{OJ2} & = -\rp{3}{2}\nk\ton{\rp{R}{a}}^2\rp{\cos I}{\ton{1-e^2}^2}, \\ \nonumber \\
\dot\Omega_{.4} \lb{OJ4} & =\dot\Om_{.2}\qua{\rp{5}{8}\ton{\rp{R}{a}}^2\rp{\ton{1 + \rp{3}{2}e^2}}{\ton{1-e^2}^2}\ton{7\sin^2 I - 4}},
\end{align}
\textcolor{black}{calculated for LAGEOS,~LAGEOS II,~LARES as in Table \ref{tavola1}}, turn out to be \cite{2005NewA...10..616I,2011Ap&SS.331..351I}
\begin{align}
c_1 \lb{ci1} &= \rp{ \Om^\textrm{LR}_{.2}~\Om^\textrm{L}_{.4} - \Om^\textrm{L}_{.2}~\Om^\textrm{LR}_{.4} }
{ \Om^\textrm{L~II}_{.2}~\Om^\textrm{LR}_{.4} - \Om^\textrm{LR}_{.2}~\Om^\textrm{L~II}_{.4} } \textcolor{black}{\simeq 0.344281069},\\ \nonumber \\
c_2 \lb{ci2} &=  \rp{ \Om_{.2}^\textrm{L}~\Om^\textrm{L~II}_{.4} - \Om_{.2}^\textrm{L~II}~\Om^\textrm{L}_{.4}}
{ \Om^\textrm{L~II}_{.2}~\Om^\textrm{LR}_{.4} -  \Om^\textrm{LR}_{.2}~\Om^\textrm{L~II}_{.4}   } \textcolor{black}{\simeq 0.073388218}.
\end{align}
In \rfrs{OJ2}{OJ4}, $R$ is the primary's equatorial radius, while $\nk=\sqrt{GM/a^3}$ and $I$ are the satellite's Keplerian mean motion and orbital inclination to the reference $\grf{x,y}$ plane, respectively; $M$ is the mass of the central body.
The combined Lense-Thirring signal, calculated from \rfr{LTtheo}, \textcolor{black}{ \rfr{comboLT}, } \rfrs{ci1}{ci2} \textcolor{black}{and \rfrs{OJ2}{OJ4}}, is
\eqi \mathcal{C}_\textrm{LT}= 50.2~\textrm{mas~yr}^{-1}.\lb{ltc}\eqf
\textcolor{black}{We point out that the numerical values of $c_1,~c_2$ in \rfrs{ci1}{ci2} are quoted with nine decimal digits  in order to assure a cancelation of $J_2$ accurate to better than $1\%$ level. If the only concern were the evaluation of the impact of the errors in the even zonals of degree $\ell\geq 6$ discussed in Section \ref{alte}, much less accurate values as $c_1\approx 0.344,~c_2\approx 0.0733$ would be adequate; nonetheless, the first even zonal harmonic would nominally impact the combined Lense-Thirring trend at a $860\%$ level. The authors of Ref.~\cite{2016EPJC...76..120C} did not quote any numerical values for $c_1,~c_2$.}

In principle, the strategy reviewed above allows one to extract also $\delta J_2$ or $\delta J_4$, provided the substitution
$\dot\Om_{.2}^{(i)}~\textrm{or}~\dot\Om_{.4}^{(i)} \rightarrow \dot\Om_\textrm{LT}^{(i)}$  is performed in \rfrs{ci1}{ci2}.

\subsection{The impact of the uncancelled even zonals of degree higher than $\ell=4$ and the need of using several Earth gravity models}\lb{alte}
A major source of systematic uncertainty, to be reliably and realistically assessed, is represented by the impact of the mismodeling in the  even zonals of degree $\ell \geq 6$ affecting the classical node precessions which are not cancelled by the adopted linear combination.

Ciufolini et al. \cite{2016EPJC...76..120C} limited themselves to use only a single Earth's global gravity field solution,  GGM05S \cite{2013AGUFM.G32A..01T}, without any motivation for such a seemingly arbitrary choice. Moreover,  in Ref.~\cite{2016EPJC...76..120C}, it is claimed that, after arbitrarily tripling the released sigmas of the even zonal coefficients of such a model, the overall systematic uncertainty would be as little as $4\%$, without providing any details on either the computational strategy adopted or the maximum degree $\ell$ considered. Furthermore, in  Ref.~\cite{2016EPJC...76..120C}, it is also claimed, without any explicit and quantitative justifications, that different Earth gravity models would allegedly lead just to slightly different results.

Actually, in view of the nowadays ever increasing number of different Earth's global gravity models released by several independent institutions worldwide\textcolor{black}{ \footnote{\textcolor{black}{See \texttt{http://icgem.gfz-potsdam.de/ICGEM/} on the Internet.}}}, a fair practice should consist of taking into account an ensemble of various models and look into the resulting scatter of the a-priori evaluations of the overall systematic uncertainties with respect to the expected Lense-Thirring trend. Moreover, correctly  assessing the realistic errors in the even zonals requires great care, following quantitative and unambiguous procedures such as those outlined in Ref.~\cite{2012JGeod..86...99W}; it should not be done in a hand-wavy fashion.

All such critical issues, ignored by the authors of Ref.~\cite{2016EPJC...76..120C}, are still valid despite the most recent advances in the field of the determination of the Earth's global gravity field models. Below, we will demonstrate it by looking at some of the most recently released geopotential models.

In general, a straightforward way to analytically calculate the long-term orbital perturbations experienced by a test particle orbiting a primary of mass $M$ due to a disturbing additional potential $U_\textrm{pert}$ with respect to the Newtonian monopole consists of taking its average over one orbital period $P_\textrm{b}$ \eqi \ang{U_\textrm{pert}}=\rp{1}{P_\textrm{b}}\int_0^{P_\textrm{b}}U_\textrm{pert} dt,\lb{ave}\eqf
and, then, inserting it in the Lagrange equations for the rates of change of the Keplerian orbital elements. In the case of the node, it is
\eqi\ang{\dert\Om t} = -\rp{1}{\nk a^2 \sqrt{1-e^2}\sin I}\derp{\ang{U_\textrm{pert}}}{I}.\lb{nodorate}\eqf
%
%
The temporal average of \rfr{ave} has to be calculated onto the unperturbed Keplerian ellipse characterized by
\begin{eqnarray}
dt &=& \rp{\ton{1-e^2}^{3/2}}{\nk \ton{1 + e\cos f}^2}df, \\ \nonumber \\
r & = & \rp{a(1-e^2)}{1 + e \cos f}, \\ \nonumber \\
x &=& r\ton{\cos\Om\cos u -\cos I\sin\Om\sin u  }, \\ \nonumber \\
y &=& r\ton{\sin\Om\cos u + \cos I\cos\Om\sin u} , \\ \nonumber \\
z &=& r\sin I\sin u,
\end{eqnarray}\lb{kep}
where $\omega,~f,~u\doteq\omega + f$ are  the argument of pericenter, the true anomaly, and the argument of latitude, respectively.

In the case of an axially symmetric primary with equatorial radius $R$ and  symmetry axis $\boldmath{\hat{k}}$, the resulting perturbing potential of degree $\ell$ and order $m=0$ is
\eqi U_{J_\ell} = \rp{GM}{r}J_\ell \ton{\rp{R}{r}}^\ell P_\ell\ton{\psi},~\psi\doteq {\boldmath{\hat{r}}}{\boldmath\cdot}{\boldmath{\hat{k}}},\lb{UL}\eqf
where $J_\ell$ is the even zonal harmonic of degree $\ell$, and $\boldmath{\hat{r}} = {\boldmath{r}}/r$ is the unit vector pointing from the primary to the test particle. If the primary's equatorial plane is assumed as reference $\grf{x,y}$ plane, as in the case of the Earth's artificial satellites, then
\eqi\psi =\sin I\sin\ton{\omega + f}.\eqf

In fact, also the uncancelled even zonals of degree as low as $\ell=6,~\textcolor{black}{8,~10}$ may have a non-negligible impact on the linear combination used because of their contribution to the nodal precession rate for LARES, as we will show in this Section.

To order zero in the eccentricity,  which is fully adequate for a quasi-circular orbit like that of LARES, \rfrs{ave}{nodorate} and \rfr{UL} straightforwardly yield
\begin{align}
\dot\Omega_{.6} \lb{Nodo6} & = -\rp{105}{1024}\nk\ton{\rp{R}{a}}^6\cos I\ton{19 + 12\cos 2I + 33\cos 4I}, \\ \nonumber \\
\dot\Omega_{.8} \lb{Nodo8} & =\rp{315}{65536}\nk\ton{\rp{R}{a}}^8\cos I\ton{178 + 869 \cos 2 I + 286 \cos 4 I + 715 \cos 6 I}, \\ \nonumber \\
\dot\Omega_{.10} \lb{Nodo10} \nonumber & = -\rp{3465}{4194304}\nk\ton{\rp{R}{a}}^{10}\cos I\ton{2773 + 2392\cos 2I + 5252 \cos 4I + \right.\\ \nonumber \\
&+\left. 1768 \cos 6I + 4199 \cos 8I }
\end{align}
for $\ell=6,~8,~\textcolor{black}{10}$. \textcolor{black}{
In the limit $e\rightarrow 0$, Equations (12) to (14) of Ref.~\cite{2003CeMDA..86..277I}, obtained with the formalism of the inclination and eccentricity functions by \cite{Kaula00}, agree with \rfrs{Nodo6}{Nodo10}. In the case of LAGEOS, LAGEOS II, LARES, \rfrs{Nodo6}{Nodo10} return the values reported in Table \ref{tavola1}.
}

On the other hand, the \textcolor{black}{remaining} scatter in the estimated values of the corresponding Stokes geopotential coefficients ${\overline{C}}_{\ell,0}$ from several global gravity models does not yet allow for an \textcolor{black}{unequivocal} and unambiguous  mismodeling \textcolor{black}{ below the $1\%$ } level in the resulting node precessions of LARES.

This can be quickly shown by considering  the simple differences $\Delta {\overline{C}}_{\ell,0}$ among the estimates for ${\overline{C}}_{\ell,0}$ for various pairs of Earth's global gravity models as a naive measure of the realistic uncertainties in the even zonals of interest. In the following, we will demonstrate that also more refined calculation yield essentially the same conclusions. \textcolor{black}{Table \ref{table1} displays the determined values ${\overline{C}}_{\ell,0}$ along with their sigmas $\sigma_{{\overline{C}}_{\ell,0}}$ of some recent field solutions for the degrees $\ell=6,~8,~10$.}
\begin{table}
\caption{Estimated values ${\overline{C}}_{\ell,0}$ and formal errors $\sigma_{ {\overline{C}}_{\ell,0} }$ of the even zonal harmonics of degree $\ell=6,~8,~\textcolor{black}{10}$ according to the Earth's global gravity field solutions GOCO05S, ITU$\_$GRACE16, ITSG-Grace2014s, JYY$\_$GOCE04S retrievable on the Internet at \texttt{http://icgem.gfz-potsdam.de/ICGEM/}. Only the significant digits are displayed. }
\label{table1}       
\begin{tabular}{lllll}
\hline\noalign{\smallskip}
 & GOCO05S & ITU$\_$GRACE16 & ITSG-Grace2014s & JYY$\_$GOCE04S  \\
\noalign{\smallskip}\hline\noalign{\smallskip}
${\overline{C}}_{6,0}$ & $-1.499663\times 10^{-7}$ & $-1.4999827\times 10^{-7}$ & $-1.499746\times 10^{-7}$ & $-1.4998\times 10^{-7}$  \\
$\sigma_{{\overline{C}}_{6,0}}$ & $1\times 10^{-13}$ & $6\times 10^{-14}$ & $2\times 10^{-13}$ & $2\times 10^{-11}$ \\
${\overline{C}}_{8,0}$ & $4.94816\times 10^{-8}$ & $4.948113\times 10^{-8}$ & $4.94779\times 10^{-8}$ & $4.938\times 10^{-8}$ \\
$\sigma_{{\overline{C}}_{8,0}}$ & $1\times 10^{-13}$ & $5\times 10^{-14}$ & $1\times 10^{-13}$ & $2\times 10^{-11}$ \\
${\overline{C}}_{10,0}$ & $5.334319\times 10^{-8}$ &  $5.335971\times 10^{-8}$ & $5.334268\times 10^{-8}$ & $5.352\times 10^{-8}$  \\
$\sigma_{{\overline{C}}_{10,0}}$ & $8\times 10^{-14}$ & $4\times 10^{-14}$ & $9\times 10^{-14}$ & $3\times 10^{-11}$ \\
\noalign{\smallskip}\hline
\end{tabular}
\end{table}
\subsubsection{\textcolor{black}{The impact of the mismodeling in the third even zonal ($\ell=6$)}}
\begin{table}
\caption{Absolute values of the differences $\Delta{\overline{C}}_{6,0}$ among the estimated values ${\overline{C}}_{6,0}$  of the models of Table \ref{table1}. They are very close to the uncertainties retrievable from the application of Eq. (A11) by Wagner and McAdoo \cite{2012JGeod..86...99W} to the sigma of the model to be tested by contrasting it to a formally superior one according to the method of such authors. }
\label{table2}       
\begin{tabular}{l|llll}
\hline\noalign{\smallskip}
 & GOCO05S & ITU$\_$GRACE16 & ITSG-Grace2014s & JYY$\_$GOCE04S  \\
\noalign{\smallskip}\hline\noalign{\smallskip}
GOCO05S & - & $3.197\times 10^{-11}$ & $8.3\times 10^{-12}$ & $1.37\times 10^{-11}$ \\
ITU$\_$GRACE16 & & - & $2.367\times 10^{-11}$ & $1.827\times 10^{-11}$ \\
ITSG-Grace2014s & & & - & $5.4\times 10^{-12}$ \\
JYY$\_$GOCE04S & & & & -\\
\noalign{\smallskip}\hline
\end{tabular}
\end{table}
It turns out that, even by restricting to $\ell=6$ alone, \textcolor{black}{whose differences $\Delta{\overline{C}}_{6,0}$ for some gravity models are displayed in Table \ref{table2}}, the corresponding mismodelled node precession of LARES $\delta\dot\Om_{J_6}^\textrm{LR}$, calculated with \textcolor{black}{the value in Table \ref{tavola1}},  \textcolor{black}{is as large as} $104~\textrm{mas~yr}^{-1}$ for the pair ITU$\_$GRACE16/GOCO05S, \textcolor{black}{while it amounts}  to $77~\textrm{mas~yr}^{-1}$ for the pair ITU$\_$GRACE16/ITSG-Grace2014s, \textcolor{black}{corresponding} to \textcolor{black}{$15\%,~11\%$}, respectively, of the combined Lense-Thirring signature of \rfr{ltc}. A smaller value can be obtained by considering ITU$\_$GRACE16 and JYY$\_$GOCE04S; indeed, it is $\delta\dot\Om_{J_6}^\textrm{LR} = \textcolor{black}{60}~\textrm{mas~yr}^{-1}$, corresponding to \textcolor{black}{$9\%$} of the predicted relativistic trend. \textcolor{black}{The complete list of percent uncertainty due to the errors in $J_6$  is quoted in Table \ref{table4}.}
\begin{table}
\caption{Percent uncertainty $\delta\mu_\textrm{LT}~(\%)$ due the errors in ${\overline{C}}_{6,0}$ according to Table \ref{table2}. }
\label{table4}       
\begin{tabular}{l|llll}
\hline\noalign{\smallskip}
 & GOCO05S & ITU$\_$GRACE16 & ITSG-Grace2014s & JYY$\_$GOCE04S  \\
\noalign{\smallskip}\hline\noalign{\smallskip}
GOCO05S & - & $15$ & $4$ & $7$ \\
ITU$\_$GRACE16 & & - & $11$ & $9$ \\
ITSG-Grace2014s & & & - & $3$ \\
JYY$\_$GOCE04S & & & & -\\
\noalign{\smallskip}\hline
\end{tabular}
\end{table}

Let us, now, adopt the more advanced approach to realistically assess the uncertainty in the even zonals proposed by Wagner and McAdoo in Ref.~\cite{2012JGeod..86...99W}.
 Eq. (A11) and Eq. (A13) of Ref.~\cite{2012JGeod..86...99W} \textcolor{black}{yield the resulting \textcolor{black}{(squared)} scale coefficients}
\begin{align}
f^2_{\textrm{test},~\ell} \lb{wag1}& = \rp{\ton{ {\overline{C}}^{\textrm{ref}}_{\ell,0} - {\overline{C}}^{\textrm{test}}_{\ell,0} }^2 - \ton{\sigma^\textrm{ref}_{ {\overline{C}}_{\ell,0} } }^2  }{\ton{\sigma^\textrm{test}_{ {\overline{C}}_{\ell,0} }}^2}, \\ \nonumber \\
g^2_{\textrm{test},~\ell} \lb{grossa} \nonumber & =
\rp{1}{2\ell + 1}\grf{ \sum_{m=0}^{\ell}\qua{   \rp{\ton{ {\overline{C}}^{\textrm{ref}}_{\ell,m} - {\overline{C}}^{\textrm{test}}_{\ell,m} }^2 - \ton{\sigma^\textrm{ref}_{ {\overline{C}}_{\ell,m} } }^2  }{\ton{\sigma^\textrm{test}_{ {\overline{C}}_{\ell,m}   }}^2}} + \right.\\ \nonumber \\
& + \left. \sum_{m = 1}^{\ell}\qua{   \rp{\ton{ {\overline{S}}^{\textrm{ref}}_{\ell,m} - {\overline{S}}^{\textrm{test}}_{\ell,m} }^2 - \ton{\sigma^\textrm{ref}_{ {\overline{S}}_{\ell,m} } }^2  }{\ton{\sigma^\textrm{test}_{ {\overline{S}}_{\ell,m}   }}^2}}}
\end{align}
\textcolor{black}{to be applied to the sigma of the Stokes coefficient of degree $\ell$ of the field labeled with \virg{test} in order to have a realistic evaluation of its uncertainty; the model labeled with \virg{ref} has a superior formal accuracy, and is used as a reference (calibrating) gravity field \cite{2012JGeod..86...99W}. }
\textcolor{black}{By adopting ITU$\_$GRACE16 as formally superior reference model and GOCO05S as test model, it turns out that the sigma of GOCO05S, quoted in Table \ref{table1}, has to be scaled by }
\begin{align}
f_{\textrm{test},~6} \lb{bi}& = 235.97, \\ \nonumber \\
g_{\textrm{test},~6} \lb{bu}& = 96.81.
\end{align}
 It can be noted that \rfr{bi} yields a realistic uncertainty for ${\overline{C}}_{6,0}$ very close to the simple difference $\Delta{\overline{C}}_{6,0}$ between the estimated coefficients of ITU$\_$GRACE16 and GGM05S.  \Rfr{bu} returns a scaling factor which is less than half the previous one; it corresponds to an overall uncertainty of $6\%$ of the combined Lense-Thirring signature.
If, instead, ITSG-Grace2014s is assumed as test model by keeping ITU$\_$GRACE16 as reference model, we have
\begin{align}
f_{\textrm{test},~6} \lb{Bi2} &= 125.65,\\ \nonumber \\
g_{\textrm{test},~6} \lb{Bu2} & = 73.21.
\end{align}
\textcolor{black}{In view of the sigma of ITSG-Grace2014s reported in Table \ref{table1},}
\textcolor{black}{ \rfrs{Bi2}{Bu2} } imply a \textcolor{black}{mismodeled} LARES node precession $\delta\dot\Om_{J_6}^\textrm{LR}$ as large as  \textcolor{black}{$78~\textrm{mas~yr}^{-1}$ and $45~\textrm{mas~yr}^{-1}$, respectively,} corresponding to an overall uncertainty in the combined Lense-Thirring trend of \textcolor{black}{$11\%$ and $7\%$, respectively}. Also in this case, \rfr{wag1} yields essentially the same result as the simple difference between the nominal coefficients of the models considered.
If the JYY$\_$GOCE04S global solution is adopted as test model, to be \textcolor{black}{compared} with  ITU$\_$GRACE16 as reference model, it turns out
that
\begin{align}
f_{\textrm{test},~6} \lb{bi2} &= 0.76,\\ \nonumber \\
g_{\textrm{test},~6} \lb{bu2} & = 1.23.
\end{align}
\textcolor{black}{By applying \rfrs{bi2}{bu2} to the sigma of JYY$\_$GOCE04S, quoted in Table \ref{table1},} the resulting bias amount to \textcolor{black}{$6\%$ and $10\%$, respectively,} of the expected gravitomagnetic signature.
\subsubsection{\textcolor{black}{The impact of the mismodeling in the fourth even zonal ($\ell=8$)}}
\begin{table}
\caption{Absolute values of the differences $\Delta{\overline{C}}_{8,0}$ among the estimated values ${\overline{C}}_{8,0}$  of the models of Table \ref{table1}. They are very close to the uncertainties retrievable from the application of Eq. (A11) by Wagner and McAdoo \cite{2012JGeod..86...99W} to the sigma of the model to be tested by contrasting it to a formally superior one according to the method of such authors. }
\label{table3}       
\begin{tabular}{l|llll}
\hline\noalign{\smallskip}
 & GOCO05S & ITU$\_$GRACE16 & ITSG-Grace2014s & JYY$\_$GOCE04S  \\
\noalign{\smallskip}\hline\noalign{\smallskip}
GOCO05S & - & $4.7\times 10^{-13}$ & $3.7\times 10^{-12}$ & $1.016\times 10^{-10}$ \\
ITU$\_$GRACE16 & & - & $3.23\times 10^{-12}$ & $1.0113\times 10^{-10}$ \\
ITSG-Grace2014s & & & - & $9.79\times 10^{-11}$ \\
JYY$\_$GOCE04S & & & & -\\
\noalign{\smallskip}\hline
\end{tabular}
\end{table}
Also the mismodeling in $\dot\Om_{J_8}^\textrm{LR}$ may give a non-negligible contribution, to be added to that due to $\ell=6$. Indeed, the differences $\Delta {\overline{C}}_{\ell,0}$ of the nominal values of the even zonal with $\ell=8$ for the pairs  ITU$\_$GRACE16/JYY$\_$GOCE04S and GOCO05S/JYY$\_$GOCE04S \textcolor{black}{listed in Table \ref{table3}, along with the precessional coefficients of Table \ref{tavola1},} yield a mismodelled precession as large as $\delta\dot\Om_{J_8}^\textrm{LR} = 40~\textrm{mas~yr}^{-1}$, corresponding to $\delta\mu_\textrm{LT} = 6\%$ of  \rfr{ltc}. \textcolor{black}{See Table \ref{table5} for the complete list of percent uncertainties $\delta\mu_\textrm{LT}$  due to all the models considered for $\ell = 8$. }
The application of the more sophisticated method by Wagner and McAdoo \cite{2012JGeod..86...99W} to the models GOCO05S (reference) and JYY$\_$GOCE04S (test) yields
\begin{align}
f_{\textrm{test},~8} \lb{Ki2} &= 4.50,\\ \nonumber \\
g_{\textrm{test},~8} \lb{Ku2} & = 1.63.
\end{align}
Such scaling factors must be applied to the sigma of JYY$\_$GOCE04S\textcolor{black}{, listed in Table \ref{table1},} providing us with an overall uncertainty in the combined relativistic effect as large as \textcolor{black}{$6\%$ and $2\%$, respectively}. \textcolor{black}{Also in this case, the scaling factor of \rfr{wag1} yields a realistic uncertainty very close to the one coming from the simple difference of the estimated values; cfr. Table \ref{table3}.}
\begin{table}
\caption{Percent uncertainty $\delta\mu_\textrm{LT}~(\%)$ due the errors in ${\overline{C}}_{8,0}$ according to Table \ref{table3}. }
\label{table5}       
\begin{tabular}{l|llll}
\hline\noalign{\smallskip}
 & GOCO05S & ITU$\_$GRACE16 & ITSG-Grace2014s & JYY$\_$GOCE04S  \\
\noalign{\smallskip}\hline\noalign{\smallskip}
GOCO05S & - & $0.02$ & $0.2$ & $6$ \\
ITU$\_$GRACE16 & & - & $0.2$ & $6$ \\
ITSG-Grace2014s & & & - & $5$ \\
JYY$\_$GOCE04S & & & & -\\
\noalign{\smallskip}\hline
\end{tabular}
\end{table}
\subsubsection{\textcolor{black}{The impact of the mismodeling in the fifth even zonal ($\ell = 10$)}}
\textcolor{black}{
From Table \ref{tavola1} and Table \ref{T10}
\begin{table}
\caption{Absolute values of the differences $\Delta{\overline{C}}_{10,0}$ among the estimated values ${\overline{C}}_{10,0}$  of the models of Table \ref{table1}. They are very close to the uncertainties retrievable from the application of Eq. (A11) by Wagner and McAdoo \cite{2012JGeod..86...99W} to the sigma of the model to be tested by contrasting it to a formally superior one according to the method of such authors. }
\label{T10}       
\begin{tabular}{l|llll}
\hline\noalign{\smallskip}
 & GOCO05S & ITU$\_$GRACE16 & ITSG-Grace2014s & JYY$\_$GOCE04S  \\
\noalign{\smallskip}\hline\noalign{\smallskip}
GOCO05S & - & $1.652\times 10^{-11}$ & $5.1\times 10^{-13}$ & $1.8\times 10^{-10}$ \\
ITU$\_$GRACE16 & & - & $1.703\times 10^{-11}$ & $1.6\times 10^{-10}$ \\
ITSG-Grace2014s & & & - & $1.8\times 10^{-10}$ \\
JYY$\_$GOCE04S & & & & -\\
\noalign{\smallskip}\hline
\end{tabular}
\end{table}
, it turns out that also the even zonal of degree $\ell = 10$ may act as a source of non-negligible systematic bias. Indeed, Table \ref{Tmu10} shows that, for the pairs GOCO05S/JYY$\_$GOCE04S, ITU$\_$GRACE16/JYY$\_$GOCE04S and ITSG-Grace2014s/JYY$\_$GOCE04S, the resulting percent uncertainties $\delta\mu_\textrm{LT}$ are as large as $\approx 30\%$.
}
\begin{table}
\caption{Percent uncertainty $\delta\mu_\textrm{LT}~(\%)$ due the errors in ${\overline{C}}_{10,0}$ according to Table \ref{T10}. }
\label{Tmu10}       
\begin{tabular}{l|llll}
\hline\noalign{\smallskip}
 & GOCO05S & ITU$\_$GRACE16 & ITSG-Grace2014s & JYY$\_$GOCE04S  \\
\noalign{\smallskip}\hline\noalign{\smallskip}
GOCO05S & - & $3$ & $0.1$ & $36$ \\
ITU$\_$GRACE16 & & - & $3$ & $32$ \\
ITSG-Grace2014s & & & - & $36$ \\
JYY$\_$GOCE04S & & & & -\\
\noalign{\smallskip}\hline
\end{tabular}
\end{table}

\textcolor{black}{The use of \rfrs{wag1}{grossa} with the ITU$\_$GRACE16 (ref) and JYY$\_$GOCE04S (test) models return
\begin{align}
f_{\textrm{test},~10} \lb{bi10}& = 5.8, \\ \nonumber \\
g_{\textrm{test},~10} \lb{qi10}& = 2.1.
\end{align}
It turns out that the  uncertainty resulting from the scaling of the sigma $\sigma_{{\overline{C}}_{10,0}}$ of JYY$\_$GOCE04S, quoted in Table \ref{tavola1},  by \rfr{bi10} is essentially identical to the difference $\Delta{\overline{C}}_{10,0}$ between the estimated values for ITU$\_$GRACE16 and JYY$\_$GOCE04S. On the other hand, using the smaller scaling factor of \rfr{qi10} returns a bias $\delta\mu_\textrm{LT}$ as large as $13\%$. }
\subsection{The determination of the corrections to $J_2,~J_4$ with the linear combination approach}
As pointed out at the end of Section \ref{methodo}, the linear combination  approach allows, in principle, to determine also the corrections $\delta J_2,~\delta J_4$ to the nominal values of the specific background reference  model adopted for the geopotential in the LAGEOSs' data reduction, or  even $J_2,~J_4$ themselves if a truncated model not including them at all is used. Such a test would be of great relevance since the results obtained in this way could be compared to those inferred in the GRACE/GOCE-based global solutions. At present, it has not been implemented.

Remarkably, Ciufolini et al. \cite{2012EPJP..127..133C} actually estimated some geopotential coefficients as solve-for parameters in a past data reduction of the LAGEOS/LARES observations. If, on the one hand, this fact further strengthens our concerns about the \textcolor{black}{continuing} lack of an explicit determination of a dedicated Lense-Thirring parameter itself, on the other hand, as remarked in Ref.~\cite{2013NewA...23...63R}, the results by Ciufolini et al. \cite{2012EPJP..127..133C}  for the estimated Stokes coefficients were up to two orders of magnitude less accurate than in the latest GRACE/GOCE-based global solutions.
\section{Summary and conclusions}
The recently published test of frame-dragging with the terrestrial geodetic satellites of the LAGEOS family by Ciufolini et al. is flawed by the same fundamental issues as many of the other previous works by the same authors.

As in previous occasions, the linear combination of the nodes of the three satellites used as observable, explicitly published by the present author  in several papers since 2005 on the basis of a method proposed by Ciufolini in 1996, was attributed also this time by Ciufolini et al. to themselves.

Our analysis shows that the uncertainties in the uncancelled even zonals of degree $\ell=6,~8,~\textcolor{black}{10}$, computed \textcolor{black}{both by taking the simple differences in the estimated geopotential's coefficients in satellite-based global models for such degrees and with a method published in the literature by Wagner and McAdoo a few years ago to quantitatively assess their realistic accuracy},  may induce a systematic bias up to \textcolor{black}{$\lesssim 15\%~(\ell=6),~6\%~(\ell=8),~36\%~(\ell=10)$ for certain models, as resumed by Table \ref{finale},}  because of the scatter in the determined values of the corresponding geopotential coefficients from several recent global gravity fields \textcolor{black}{produced by different institutions worldwide}.
\begin{table}
\caption{Largest values of the percent uncertainties $\delta\mu_\textrm{LT}~(\%)$ by degree $\ell$ according to Tables \ref{table4}, Table \ref{table5}, and Table \ref{Tmu10}.  }
\label{finale}       
\begin{tabular}{llll}
\hline\noalign{\smallskip}
& $\ell=6$ & $\ell=8$  & $\ell=10$  \\
\noalign{\smallskip}\hline\noalign{\smallskip}
$\delta\mu_\textrm{LT}$ & 15 (GOCO05S/ITU$\_$GRACE16) & 6 (GOCO05S/JYY$\_$GOCE04S) & 36 (GOCO05S/JYY$\_$GOCE04S)\\
\noalign{\smallskip}\hline
\end{tabular}
\end{table}

As a further consistency test of the approach followed so far, we suggest to use suitably designed linear combinations to determine also the corrections \textcolor{black}{ $\delta J_2,~\delta J_4$ } to the reference background geopotential model used in the reduction of the LAGEOSs data.

Finally, it is  remarkable that, after about twenty years since the first reported tests with LAGEOS and LAGEOS II and four years since the launch of LARES, nobody has yet published any genuinely independent test of the Lense-Thirring effect with such geodetic satellites in the peer-reviewed literature, especially in view of how many researchers around the world constitute the global satellite laser ranging community.
\begin{acknowledgements}
I am grateful to an anonymous referee for her/his steady efforts to improve my manuscript with constructive and careful remarks.
\end{acknowledgements}

\bibliographystyle{spmpsci}      
\bibliography{Gclockbib}   

\end{document}